\documentstyle[preprint,eqsecnum,aps,epsf]{revtex}
\begin{document}
\tolerance=10000
\hfuzz=5 pt
\baselineskip=24 pt
\draft
\preprint{HD-THEP-97-60}

\title{\bf Field Strength Correlators and Dual Effective Dynamics
 in QCD}
\author {M. Baker}
\address{University of Washington, Seattle, WA 98105, USA}
\author{N. Brambilla $^*$, H. G. Dosch, 
 A. Vairo\thanks{Alexander von Humboldt Fellow} }
\address{\it Institut f\"ur Theoretische Physik, Universit\"at Heidelberg\\ 
 Philosophenweg 16, D-69120 Heidelberg, FRG}
\maketitle

\begin{abstract}
\baselineskip=20 pt 
\noindent 
We establish a relation between the two-point field strength correlator 
in QCD and the dual field propagator of an effective dual Abelian Higgs model 
describing the infrared behaviour of QCD. We find an analytic approximation 
to the dual field propagator without sources and in presence of quark sources. 
In the latter situation we also obtain an expression for
the static $q \bar{q}$ potential. Our derivation sheds some light 
on the dominance and phenomenological relevance of the two-point field 
strength correlator. 
\end{abstract}


\vfill
\eject
\section{INTRODUCTION}

The gauge invariant field strength vacuum correlators
\begin{equation}
\langle F_{\mu_1\nu_1} (x_1) U(x_1,x_2) F_{\mu_2\nu_2} (x_2) U(x_2,x_3) \cdots 
F_{\mu_n\nu_n}(x_n)U(x_n,x_1)\rangle,
\label{cor}
\end{equation}
where $U(x,y) \equiv {\rm P} \displaystyle\exp ig \int_y^x dz^\mu A_\mu(z)$   
is the Schwinger color string, play a relevant role in gluodynamics 
with and without quark sources. We know that in the infrared region 
these correlators are dominated by their non-perturbative behaviour. 
In particular  the non-perturbative ``gluon condensate''  
\begin{equation}
\langle {\alpha_s\over \pi }  F_{\mu \nu}^a(0) 
F^{\mu \nu}_a(0)\rangle^{\rm non \, pert}\equiv F_2,
\label{cond}
\end{equation} 
plays a crucial role  in the QCD sum rule method \cite{svz}.

The non-perturbative part of the gauge invariant two-point field 
strength correlator $\langle F_{\mu \nu}(x)U(x,0) F_{\rho \sigma}(0) U(0,x)\rangle$  
has been calculated  on the lattice, with the 
cooling method in \cite{digiaco} and in presence of sources in \cite{banoi}. 
We define (in Euclidean space-time, as in the rest of this work)
the gauge invariant correlator \cite{svm} 
\begin{eqnarray}
\langle g^2 F_{\mu\nu}(x) U(x,0) F_{\lambda\rho}(0)U(0,x) \rangle &\equiv&  
(\delta_{\mu\lambda}\delta_{\nu\rho} - \delta_{\mu\rho}\delta_{\nu\lambda}) 
g^2 D(x^2) + {1\over 2}\left[
{\partial\over\partial x_\mu}\left( x_\lambda \delta_{\nu\rho} 
- x_\rho \delta_{\nu\lambda} \right) \right.
\nonumber\\
&+&\left.  {\partial\over\partial x_\nu}
\left( x_\rho \delta_{\mu\lambda} - x_\lambda \delta_{\mu\rho} \right) \right] 
g^2 D_1 (x^2) \, . 
\label{decom}
\end{eqnarray}
A parameterization of the form 
\begin{eqnarray}
&~& D(x^2) = A e^{-|x|/T_g} + {a\over x^4}e^{-|x|/T_p}
\qquad D_1(x^2) = B e^{-|x|/T_g} + {b\over x^4}e^{-|x|/T_p}
\label{para}\\
&~& A \simeq 128 ~{\rm GeV}^4,  \qquad B  \simeq 27 ~{\rm GeV}^4,   
\qquad a \simeq 0.69, \qquad b \simeq 0.46, 
\nonumber\\
&~& T_g \simeq 0.22 ~{\rm fm}, \qquad T_p \simeq 0.42 ~{\rm fm}, 
\nonumber 
\end{eqnarray}
yields a very good fit to the (cooled) lattice data \cite{digiaco} 
in the range $ 0.1 \, {\rm fm} \leq x \leq 1 \, {\rm fm}$.
At short distances the $1/x^4$ term, which is of perturbative origin, dominates,
while at distances $x \geq 0.4 \,{\rm fm}$ the non-perturbative term, 
proportional to $e^{-{x/T_g}}$, becomes more important.
In an Abelian theory without monopoles the Bianchi identities yields $D=0$.

In the Stochastic Vacuum Model (SVM) \cite{svm,svm2,svmrev,conf} it is assumed
that for processes which can be reduced to the calculation of Wilson loops 
with quasi-static sources (such as heavy quark potentials and soft 
high energy scattering amplitudes in the eikonal approximation) 
the infrared behaviour of QCD can be approximated by a Gaussian stochastic 
process in the field strength and is thus determined approximately
by the correlator (\ref{decom}). In particular also the Wilson loop average 
is given only in terms of (\ref{decom}).

We know from the strong coupling expansion and lattice simulations that the Wilson 
loop is an order parameter for confinement. The confining area law behaviour of
the Wilson loop  is reproduced by the stochastic vacuum model provided that 
the form factor $D$ is different from zero and is dominated in the infrared  
region by a decreasing behaviour with the fall off controlled by a finite 
correlation length $T_g$. These features of $D$ are compatible with lattice data 
(see (\ref{para})). Furthermore this model gives a good description of
certain features of high energy scattering (e. g. \cite{doschscat}). We will come back 
to this point in Sec. 2.

It is the goal of this paper to relate the gluon correlator to  
the Mandelstam--'t Hooft dual superconductor mechanism of confinement \cite{man}.
In this picture the physical essence of the confinement 
is the formation of color-electric flux tubes between quarks 
due to a dual Meissner effect. The monopoles condense
and lead to a dual superconductor which forces the color-electric field lines 
in flux tubes which are the dual analogue to the Abrikosov--Olesen strings. 
The formation of an electric flux tube is also the consequence
of the Stochastic Vacuum Model \cite{rued}.

Furthermore, in an Abelian projection of QCD, monopoles
are the degrees of freedom responsible for confinement.
Monopoles condensation has been observed on the lattice 
(for a review see \cite{pol}) and when confinement can be derived 
analytically  (compact electrodynamics, Georgi--Glashow model, 
and some supersymmetric Yang--Mills theories), it  is due 
to the condensation of monopoles. The monopole potential can 
be measured in the Abelian projection and it turns out that 
in the confining phase it has the Higgs form \cite{pol2}. 
Lattice measurements of the distribution of monopole currents indicate that 
at large distances gluodynamics is equivalent  to a dual Abelian Higgs model, 
the Higgs particles are Abelian monopoles and these are condensed in the 
confining phase.  In the maximal Abelian gauge the structure of the 
interquark flux tube was intensively studied and high precision measurements 
of the colour fields and the monopole currents recently allowed for 
a detailed check of the dual superconductor scenario with respect 
to the Ginzburg--Landau equations \cite{bal1}. 

Analytic  models of the infrared dynamics of dual QCD with monopoles  were 
constructed  \cite{baker1,suzuki} and their phenomenological  consequences
intensively investigated. In the effective 
dual model of Baker, Ball and Zachariasen (dual QCD, DQCD),  
the  complete semirelativistic  quark-antiquark potential, the 
flux tube distribution and the energy density  were obtained 
from the numerical solution of the coupled non-linear equations of motion 
and compared very favourably with recent lattice data  
\cite{bakerlat,bakerflux,green}. Although the Lagrangian of this 
effective dual theory for long distance QCD is based on a non-Abelian 
gauge group, the results for the $q \bar{q}$ potentials 
aside from an overall color factor  can to a very good approximation 
be described by a (dual) Abelian Higgs model. Therefore, the results are in this case 
largely insensitive to the details of the dual gauge group and the quarks select 
out only Abelian configurations of the dual potential \cite{BBBZ}.  

Since an  effective  Abelian description of the infrared confining dynamics 
of QCD (at least for heavy quarks) emerges either from QCD 
(via Gaussian approximation and bilocal strength tensor 
correlators\footnote{In the treatment of two Wilson loops, however, 
the non-Abelian characteristics of QCD becomes very important 
\cite{doschscat,rued}.}) or via an effective Abelian Higgs models 
it becomes extremely interesting to exploit in which sense the two Abelian 
descriptions  are equivalent and, once we assume an equivalence, what 
kind of constraints this imposes on the form of the QCD field strength 
correlators. In the present work we will obtain from the dual Abelian 
Higgs model information on the form of the gauge invariant two-point field 
strength correlator (\ref{decom}) and in addition we will obtain an analytic 
approximation for the static heavy quark potential given by the dual theory.

There are  two more arguments which motivate such an investigation. 
First, as we will discuss briefly in Sec. 2, the existence of a non-vanishing 
form factor $D$ in the two-point field strength correlator of QCD 
seems to suggest quite naturally the existence of an effective free dual 
Abelian theory ``behind'' the long-range dynamics of QCD.
Second, a recent comparison between the complete semirelativistic 
potentials obtained in DQCD and in the Gaussian stochastic 
approximation of QCD in the limit of large interquark distances 
showed up quite striking and surprising similarities (see \cite{bvpot}). 
The following analysis wants to shed some light on that. 

The plan of the paper is the following. In Sec. 2 we recollect some essential 
features of the gauge invariant two-point correlator in QCD and we establish 
a relation with the Wilson loop. In Sec. 3 we investigate the analogous quantity 
in the dual Abelian Higgs model without sources.  For a constant Higgs field we 
reproduce a two-point correlator having the same behaviour as
obtained by other authors by studying the London limit of a dual 
Abelian Higgs model. In Sec. 4 we introduce sources and obtain 
an analytic expression for the static potential.  This suggests a connection 
between the parameters of the two-point field strength correlator in QCD 
and those of the dual Abelian Higgs model. Finally, Sec. 5 contains some conclusions.

\section{GAUGE-INVARIANT TWO-POINT GLUON CORRELATOR AND WILSON LOOP IN QCD}

We consider the correlator of two gluon field strengths in QCD at different 
space-time points, connected by a Schwinger string. This string can either 
consist of two strings in the fundamental representation  or one string in the adjoint 
one. For definiteness in notation we choose the first possibility 
and consider the quantity 
$$
\langle g^2 F_{\mu\nu}(x)\, U(x,0) F_{\lambda\rho}(0) U(0,x) \rangle .
$$
The Lorentz decomposition of this correlator is given by Eq. (\ref{decom})
and the results of the lattice measurements are collected in Eq. (\ref{para}).
The leading (tree level) perturbative contribution is contained in the 
form factor $D_1$. In an Abelian gauge theory without 
monopoles the Bianchi identity implies  that the form factor $D$
vanishes identically \cite{svm}. In a non-Abelian theory or in an Abelian 
theory with monopoles $D$ can be different from zero. Let us briefly 
review how a non-vanishing $D$ leads to confinement \cite{svm}. 

In the presence of heavy quark sources the relevant object in QCD is the Wilson loop 
average $W(\Gamma)$, where $\Gamma$ is a closed curve built up by the 
trajectories of external sources  and some Schwinger strings connecting the end-points. 
By means of the non-Abelian Stokes theorem \cite{stokes} one can express the Wilson 
loop average $W(\Gamma)$ in terms of an integral over a surface 
$S(\Gamma)$ enclosed by the contour $\Gamma$. A way to evaluate analytically 
this quantity consists in expanding this expression via a cluster expansion and 
 keeping only the bilocal cluster (i. e. in assuming that the vacuum 
fluctuations are of a Gaussian type)\footnote{For an extensive discussion 
on the validity of this assumption see \cite{conf}. Moreover, recent lattice 
calculations  seem to confirm that heavy quark potentials are really 
dominated by the two-point gluon field strength correlator \cite{banoi}.}:
\begin{eqnarray}
W(\Gamma) &\equiv& \left\langle {\rm P} \>
\exp \left( ig \int_{\Gamma} dz_\mu  A_\mu(z) \right) \right\rangle 
{\mathop{=}\limits_{{\rm Stokes}}}
\left\langle {\rm P} \> \exp \left( ig \int_{S(\Gamma)} dS_{\mu\nu}(u) 
F_{\mu\nu}(u,x_0) \right) \right\rangle 
\label{stokes}\\ 
&{\mathop{\simeq}\limits_{{\rm SVM}}}& \exp\left(-{1\over 2} 
\int_{S(\Gamma)} dS_{\mu\nu}(u) \int_{S(\Gamma)} dS_{\lambda\rho} (v)  
\langle g^2 F_{\mu\nu}(u,x_0)  F_{\lambda\rho}(v,x_0) \rangle\right), 
\label{svm}
\end{eqnarray}
where  ${\rm P \,}F_{\mu\nu}(u,x_0) \equiv {\rm P \,} U(u,x_0) F_{\mu\nu}(u) U(x_0,u)$.
Assumption (\ref{svm}) corresponds to the so-called Stochastic Vacuum 
Model (SVM) \cite{svm}. The point $x_0$ is an arbitrary reference point on the surface 
$S(\Gamma)$ needed  for surface ordering. Of course the final result in 
the full theory  does not depend on the reference point $x_0$. 
The results obtained in the Gaussian approximation, however, will generally 
depend on it. This  dependence is minimized by choosing $S(\Gamma)$ 
to be the minimal area surface with contour $\Gamma$ \cite{Si}.  
Under this condition one may  neglect the $x_0$ dependence on 
$\langle g^2 F_{\mu\nu}(u,x_0)  F_{\lambda\rho}(v,x_0) \rangle$ and recover  
in this way translational invariance. Then, the decomposition 
of Eq. (\ref{decom}) can be used (by replacing $x^2$ with $(u-v)^2$). 

All the  spin and velocity dependent potentials  up to order $1/m^2$ 
in the quark mass can be expressed in terms of the functions 
$D$ and $D_1$ \cite{svm2,bvpot,doschspin}. In particular the  
static potential is given by 
\begin{equation}
V_0(R) =  {g^2\over 2} \int_{|x_1|<R} d^2x ~(R-|x_1|)~D(x^2) 
+ {|x_1|\over 2}D_1(x^2), 
\label{v0svm}
\end{equation}
with  $d^2x = dx_1 dx_4$, 
$x^2 = x_1^2 + x_4^2$. 
The string tension emerges for large $q\bar{q}$ distances $R$ as
\begin{equation}
\sigma = {g^2 \over 2} \int d^2x  \,D(x^2).
\label{sigma}
\end{equation}
Therefore  a non-vanishing  $D$ function leads  
to confinement.\footnote{The $1/x^4$ term in D in (\ref{para}) is a one-loop 
perturbative contribution \cite{jamin} and has not to be considered in the calculation 
of the string tension $\sigma$.  Preliminary results indicate that these perturbative 
contributions to $D$ appearing at one loop and higher orders are cancelled 
by higher order correlator contributions \cite{simpriv}. This is not surprising 
since in a non-Abelian theory perturbative contributions beyond the tree level 
are surely not of a Gaussian type.}

While lattice data confirm the existence of a non-vanishing form factor $D$ 
with exponential fall off, up to now there is no analytic tool which allows 
to calculate and to interpret the non-perturbative contributions to $D$ in the 
long-range regime. 

We observe, however, that a non-vanishing function $D$ emerges naturally 
if we assume that there exists an effective ``dual'' Lagrangian 
describing an Abelian gauge theory for which the dual two-point field strength 
correlator coincides in the long-range limit with the bilocal cumulant given 
by Eq. (\ref{decom}). Let us call  $G_{\mu\nu}$ the (Abelian) field 
strength of the dual theory. Since we assume this theory to 
observe the Bianchi identities we have in general
$$
\langle g^2 G_{\mu\nu}(x) G_{\lambda\rho}(0) \rangle \equiv 
{1\over 2} \left[
{\partial\over\partial x_\mu}\left( x_\lambda \delta_{\nu\rho} 
- x_\rho \delta_{\nu\lambda} \right)  
+ {\partial\over\partial x_\nu}\left( x_\rho \delta_{\mu\lambda} 
- x_\lambda \delta_{\mu\rho} \right) \right] 
g^2 \Delta (x^2).
$$ 
The expectation value of the dual of the dual fields, 
$\tilde{G}_{\mu\nu} \equiv \displaystyle{1\over 2} 
\epsilon_{\mu\nu\alpha\beta}G_{\alpha\beta}$, is 
\begin{eqnarray}
\langle g^2 \tilde{G}_{\mu\nu}(x) \tilde{G}_{\lambda\rho}(0) \rangle 
&=& (\delta_{\mu\lambda}\delta_{\nu\rho} - \delta_{\mu\rho}\delta_{\nu\lambda})
g^2 \left(\Delta(x^2) -x^2 {d \over d x^2} \Delta(x^2)\right)
\nonumber\\ 
&+& 2 \left[
{\partial\over\partial x_\mu}\left( x_\lambda \delta_{\nu\rho} 
- x_\rho \delta_{\nu\lambda} \right) 
+ {\partial\over\partial x_\nu}\left( x_\rho \delta_{\mu\lambda} 
- x_\lambda \delta_{\mu\rho} \right) \right] g^2 
{d \over dx^2}{d \over dx^2} \Delta(x^2).  
\nonumber
\end{eqnarray} 
It shows a tensor structure like the one multiplying $D$ in equation (\ref{decom}). 
The existence of such a correlator therefore seems to suggest the existence of 
a dual Abelian gauge theory for which at big distances the field strength 
correlator behaves as the corresponding correlator  of the dual theory:
\begin{equation}
\langle g^2 F_{\mu\nu}(x,x_0) F_{\lambda\rho}(0,x_0) \rangle  \sim 
\langle g^2 \tilde{G}_{\mu\nu}(x) \tilde{G}_{\lambda\rho}(0) \rangle. 
\label{naiv}
\end{equation}

In  the next section we want to explore some consequences of 
Eq. (\ref{naiv}). In Sec. 4 Eq. (\ref{naiv}) will be replaced 
by a better  founded assumption on the Wilson loop. 
Nevertheless the basic idea behind (\ref{naiv}) will remain. 

\section{DUAL ABELIAN HIGGS MODEL WITHOUT SOURCES OR VORTICES}

The aim of  this section is  essentially pedagogical. 
We will reproduce in a clear and economical way 
some of the results which can be found in the existing 
literature on the London limit of a dual Abelian Higgs model. 
We will prove in this way that assumption (\ref{naiv})  
is reasonable, i. e. compatible with (\ref{para}). 
We will also show the drawbacks of this approach and try 
to justify why we need to take into account external charge sources. 
This will lead to the results of Sec. 4.

Let us consider a very naive context, i. e. a ``dual'' vector 
gauge field $C_\mu$ minimally coupled with some external scalar 
field $\phi$ which we could call a Higgs field. The action is given by
\begin{equation}
S(C_\mu,\phi) = \int d^4x \left[ {1\over 4} G_{\mu\nu}(x) G_{\mu\nu}(x) 
+ {1\over 2}(D_\mu \phi)^*(x) (D_\mu \phi)(x) + V(\phi^*(x) \phi(x)) \right], 
\label{action}
\end{equation}
where $G_{\mu\nu}(x) = \partial_\mu C_\nu(x) - \partial_\nu C_\mu(x)$ and 
$V(\phi^*\phi) = \displaystyle{\lambda \over 4} (\phi^*\phi -\phi_0^2)^2$ 
(with $\phi_0$ different from zero). The Higgs field 
is coupled to the gauge field $C_\mu$ via the covariant derivative 
$D_\mu \phi = (\partial_\mu + i e C_\mu)  \phi$. 
  
We choose a gauge in which the regular part of the phase of $\phi$ 
vanishes, the so-called unitary gauge.
The propagator, ${\cal K}_{\mu\nu} \equiv \langle C_\mu C_\nu \rangle$,  
of the field $C_\mu$ satisfies the equation: 
\begin{equation}
(\partial^2 \delta_{\nu\mu} - \partial_\nu \partial_\mu 
-   e^2 \phi^2(x) \delta_{\nu\mu}) {\cal K}_{\nu\alpha}(x,y) 
= - \delta_{\mu\alpha}\delta^4(x-y). 
\label{KK}
\end{equation}
The quantity in which we are interested is what we could call the ``dual'' of 
the field strength two-point correlator in the theory described by the action 
(\ref{action}): 
\begin{equation}
{\cal G}_{\sigma\gamma\lambda\rho}(x,y) \equiv
(\delta_{\lambda\sigma}\delta_{\rho\gamma} 
- \delta_{\lambda\gamma}\delta_{\rho\sigma}) \delta^4(x-y) 
- \epsilon_{\mu\nu\lambda\rho}\epsilon_{\beta\alpha\sigma\gamma} 
\partial^y_\beta \partial^x_\mu {\cal K}_{\nu\alpha}(x,y). 
\label{GG}
\end{equation}
 For a matter of convenience we 
prefer to define ${\cal G}_{\sigma\gamma\lambda\rho}$ with the delta contribution 
subtracted out explicitly. 
In this model $ {\cal G}_{\sigma\gamma\lambda\rho}$ is the equivalent of 
the quantity $ \langle g^2 \tilde{G}_{\mu\nu}(x) \tilde{G}_{\lambda\rho}(y)\rangle $
introduced at the end of the last Section.  Eqs. (\ref{GG}) and (\ref{naiv}) then 
 give the correlator (\ref{decom}) in terms of the propagator of the dual theory.

Let us study, now, the case where the Higgs field has the constant value $\phi_0$. 
Then, Eq. (\ref{KK}) can be written  as:
\begin{equation}
(\partial^2 -  e^2 \phi_0^2) {\cal K}^\infty_{\mu\alpha}(x,y) = 
- \left( \delta_{\mu\alpha} 
- {\partial_\mu\partial_\alpha\over  e^2 \phi_0^2} \right) \delta^4(x-y). 
\label{KKoo}
\end{equation}
This is simply the equation defining the free propagator of 
a massive vector boson with mass $M \equiv  e \phi_0$:
\begin{eqnarray}
{\cal K}^\infty_{\mu\alpha}(x,y) &=& 
\left( \delta_{\mu\alpha} - {\partial_\mu\partial_\alpha\over M^2} \right) 
{\cal K}^\infty(x-y), \nonumber\\
\hbox{with}\> {\cal K}^\infty(x-y) &=& \int {d^4 p\over (2\pi)^4} 
e^{-ip(x-y)} {1 \over p^2 + M^2} = {M\over (2\pi)^2}{ K_1(M x) \over x}, 
\nonumber
\end{eqnarray}
where $K_n$ ($n=0,1,2,\dots$) are Bessel functions. As a consequence we can write 
\begin{eqnarray}
{\cal G}_{\sigma\gamma\lambda\rho}(x-y) &=& 
(\delta_{\lambda\sigma}\delta_{\rho\gamma} - 
\delta_{\lambda\gamma}\delta_{\rho\sigma}) D^\infty((x-y)^2)  
+ {1\over 2}\left[
{\partial\over\partial x_\lambda}\left( (x-y)_\sigma \delta_{\rho\gamma}  
- (x-y)_\gamma \delta_{\rho\sigma} \right) \right.
\nonumber\\
&+& \left. {\partial\over\partial x_\rho}\left( (x-y)_\gamma 
\delta_{\lambda\sigma} - (x-y)_\sigma \delta_{\lambda\gamma} \right) 
\right] D^\infty_1((x-y)^2),
\label{GGoo}
\end{eqnarray}
with 
\begin{eqnarray}
D^\infty(x^2) &=& \delta^4(x) + \partial^2 {\cal K}^\infty(x^2) =  
M^2  {\cal K}^\infty(x^2) = {M^3\over 4\pi^2} {K_1(M x)\over x},
\label{Doo}\\
D^\infty_1(x^2) &=& -4 {d\over dx^2} {\cal K}^\infty(x^2) = 
{M\over 2 \pi^2 x^2}\left[{K_1(Mx)\over x}+{M\over 2}(K_0(Mx)+K_2(Mx))\right]. 
\label{D1oo}
\end{eqnarray} 
Therefore, the assumption that ${\cal G}_{\sigma\gamma\lambda\rho}$ has the 
same long-range behaviour of the gauge invariant two-point field strength 
correlator in QCD (see Eq. (\ref{naiv})) is compatible with 
the parameterization (\ref{para}) and leads to a correlation length $T_g$ 
equal to the inverse of the dual gluon mass $M$. 
In particular, the asymptotic behaviours of ${\cal K}^\infty$ are:
\begin{eqnarray}
{\cal K}^\infty(x^2) &{\mathop{\longrightarrow}\limits_{|x| \to 0}}& 
{1\over (2\pi)^2}{1\over x^2} + \cdots ,
\label{Koozero}\\
{\cal K}^\infty(x^2) &{\mathop{\longrightarrow}\limits_{|x| \to \infty}}& 
{1\over 2}{1\over (2\pi)^{3/2}}{1\over \sqrt{M} x^{3/2}} e^{-Mx} + \cdots .
\label{Kooinfty}
\end{eqnarray}

The results shown here coincide with those obtained from the 
London limit of a dual Abelian Higgs model in \cite{antonov}. 
The seeming difference for what concerns the function $D$ is due 
to the fact that we have subtracted out explicitly 
in our definition of ${\cal G}_{\sigma\gamma\lambda\rho}$ the delta 
singularity which in the referred work is taken into account in 
a regularized form. One may wonder how the result of a topologically trivial model 
(no singular Higgs phase) agrees with results which take into account 
properly the internal Abrikosov--Nielsen--Olesen strings. This is due to the fact that   
${\cal G}_{\sigma\gamma\lambda\rho}$ is sensitive to the string  
only via the strength of  the Higgs field and this is fixed to a constant here 
as well as in the London limit of a dual Abelian Higgs model. 

The agreement between both approaches reveals a common weakness: 
the missing treatment of the interaction between the internal strings 
present in the dual Abelian Higgs model and the string between external 
quark sources. In Ref. \cite{antonov} no external sources were introduced 
and the result (\ref{Doo})-(\ref{D1oo}) for the correlator 
was obtained in the following way. The functional integral 
for the Abelian Higgs model was rewritten in such a form as to exhibit 
integration  over the closed surfaces of the (internal) strings. 
From the form of the contribution of a single closed surface in the 
London limit it was deduced that it could be obtained  by a correlator 
like (\ref{Doo})-(\ref{D1oo}) in the Gaussian approximation. 
Implicitly this form was assumed to be valid also for external sources. 
When external quark sources are introduced however, the strings of 
those sources will interfere with the internal strings.  
Some aspects of that phenomenon (the linking number) have been treated 
in \cite{pol3}, but to our knowledge  there exists no analytic attempt 
to evaluate the influence on the phenomenologically relevant parameters.

We notice that, due to the short-range behaviour (\ref{Koozero}), 
Eq. (\ref{D1oo}) reproduces the expected short-range behaviour 
of the function $D_1$ ($\sim 1/x^4$, see (\ref{para})). 
Due to the short distance behaviour of the function $D$, like $1/x^2$, 
the string tension we obtained using  (\ref{naiv}), 
 Eq. (\ref{Doo}) and  Eq. (\ref{sigma}), is logarithmically divergent: 
\begin{equation}
\sigma^{\infty} \equiv {g^2 \over 2}
{\mathop{\lim}\limits_{\epsilon \to 0}} \int_{|x|>\epsilon} d^2x \,  D^\infty (x^2) 
= \pi \phi_0^2 \,\, {\mathop{\lim}\limits_{\epsilon \to 0}}\, 
K_0(\epsilon) \sim \pi\phi_0^2 \,(\log 2-\log\epsilon-\gamma),  
\label{sigmaoo}
\end{equation}
where we have used the Dirac quantization condition $e = 2\pi/g$,
relating $g$ to the coupling constant $e$ of the dual theory. 
The divergence is a short-distance effect and appears to be a result 
of the freezing of the Higgs field to the vacuum value $\phi_0$, i. e.,  
in terms of the dual Abelian Higgs model, of the London limit. 
Assuming a coordinate dependent Higgs mass  going to zero like $|x|$ near 
the origin, would yield  a finite short-range behaviour of the function $D^\infty$ 
while  preserving the perturbative short-range behaviour of the function 
$D_1^\infty$. There is, however, no motivation for such an anisotropic 
behaviour of the Higgs field unless we introduce 
some charges into the vacuum. Only in such a context we can expect 
that near the sources and on the connecting flux tube string the 
Higgs field vanishes while far away it assumes the vacuum value $\phi_0$.  
This will be precisely the subject of the next section, 
where we will consider a dual Abelian Higgs model with external charges and 
where we will also change our intuitive duality assumption (\ref{naiv}) 
to a more physically justified one. Moreover, we recall here that recent lattice 
data \cite{bal1} confirm that in the presence of external quark sources the distribution 
of electric fields and monopoles currents does not fulfill the London limit. 

To conclude this section  we  comment briefly 
on the translational invariance of the considered correlators. 
As long as $\phi$ is considered as an external 
field in Eq. (\ref{KK}), ${\cal G}_{\sigma\gamma\lambda\rho}$ is not translational 
invariant and therefore in order to take advantage of
the decomposition (\ref{decom}) we have to fix our reference frame in 
such a way that the point $y$ coincides with the origin. 
This fact is by itself not in contradiction with the duality 
assumption (\ref{naiv}) since also the correlator in 
the direct theory, $\langle F_{\mu\nu}(x,x_0) F_{\lambda\rho}(y,x_0) \rangle$, 
is in general not translational invariant, and only by choosing 
the reference point $x_0$ on the straight line connecting $x$ with $y$ 
is invariance  recovered. Finally, we notice that ${\cal G}_{\sigma\gamma\lambda\rho}$ 
is translational invariant in some particular cases: if we assume $\phi$ 
constant, as we have done in this section, or partially (in the longitudinal 
coordinates) if we assume that $\phi$  depends only  on some (transverse) 
coordinates. This last situation will be exploited in the next section.

\section{DUAL ABELIAN HIGGS MODEL WITH EXTERNAL QUARK SOURCES}

In this section, for the reasons stated above, we want to consider 
a dual Abelian Higgs model with external quark sources. In particular we want 
to make a duality assumption on the long-range behaviour of the Wilson loop 
associated with the dynamics of a two heavy quark bound state. This assumption 
will take the place of our previous statement (\ref{naiv}). We will 
see that some general features will, nevertheless, be preserved. 

Following \cite{BBBZ} we assume that the long-range behaviour of the Wilson loop 
average $W(\Gamma)$ associated with a two heavy quark bound state 
is described by the functional generator of a dual Abelian 
Higgs model with external quark sources :
\begin{equation}
W(\Gamma) \sim \langle e^{-S(C_\mu,\phi)} \rangle , 
\label{main}
\end{equation}
where the bracket $\langle ~~ \rangle$ means the average over the gauge 
fields $C_\mu$ and the Higgs field $\phi$. The Abelian Higgs model is dual 
in the sense that it is weakly coupled. Therefore the right-hand side of 
Eq. (\ref{main}) can be evaluated via a classical expansion.

The action  $S$ is given by equation (\ref{action}), but 
since we want that in the  $\lambda = 0$ limit  $S$  
describes the dual of a $U(1)$ Yang--Mills theory with two external 
point-like charge sources $-g$ (particle) and $g$ (antiparticle), we define 
the field strength tensor $G_{\mu\nu}$, now, in such a way that it 
contains not only the dual gauge fields $C_\mu$ but also
the field of the external 
sources \cite{bak}:
\begin{equation}
G_{\mu\nu}(x) = \partial_\mu C_\nu(x) - \partial_\nu C_\mu(x) 
+ G^{\rm S}_{\mu\nu}(x), 
\label{gmn}
\end{equation}
where 
\begin{equation}
G^{\rm S}_{\mu\nu}(x) = g \epsilon_{\mu\nu\alpha\beta} 
\int_0^1 d\tau \int_0^1 d\sigma {\partial y_\alpha \over \partial \sigma} 
{\partial y_\beta \over \partial \tau} \delta^4(x-y(\tau,\sigma)),
\label{dirac}
\end{equation}
and $y_\mu (\tau,\sigma)$ is a parameterization of a surface 
$S(\Gamma)$ swept by the Dirac string connecting the charges 
$-g$ and $g$. Therefore $S(\Gamma)$ is a surface with a fixed contour 
given by $\Gamma$ ($y_\mu(\tau,1) = z_{1\,\mu}$ and 
$y_\mu(\tau,0) = z_{2\,\mu}$, where  $z_{1\,\mu}$ and $z_{2\,\mu}$ are the 
charge source trajectories). Notice that the divergence of the dual of 
$G^{\rm S}_{\mu\nu}$ is just the current carried by a charge $g$ moving 
along the path $\Gamma$: $\partial_\beta \tilde{G}_{\alpha\beta}^{\rm S}(x) 
= - g \displaystyle{\oint_\Gamma dz_\alpha \delta^4(x-z)}$. The charge $g$ 
is related to $e$ by the usual Dirac quantization condition $e = 2\pi/g$. 

The leading long distance approximation to the dual theory is the classical 
approximation
\begin{equation}
\langle e^{-S(C,\phi)} \rangle \sim e^{-S(C_\mu^{\rm cl},\phi^{\rm cl})}.
\label{clas}
\end{equation}
where $C_\mu^{\rm cl}$ and $\phi^{\rm cl}$ are solutions of the equations of motion:
\begin{eqnarray}
(\partial^2 \delta_{\nu\mu} - \partial_\nu \partial_\mu 
-  e^2 \phi^2(x) \delta_{\nu\mu}) C_\nu(x) &=& 
-\partial_\nu G_{\nu\mu}^{\rm S}(x), 
\label{c2}\\
(\partial_\mu + i e C_\mu(x)) (\partial_\mu + i e C_\mu(x))\phi(x) 
&=&  \lambda (\phi^2(x) - \phi_0^2)\phi(x) .
\label{phi2}
\end{eqnarray} 
Using these equations it is possible to write 
$S(C_\mu^{\rm cl},\phi^{\rm cl})$ as:
\begin{eqnarray}
S(C_\mu^{\rm cl},\phi^{\rm cl}) &=& 
\int d^4x \int d^4y {1\over 2} G^{\rm S}_{\beta\alpha}(y) 
\left[ {1\over2} \delta_{\beta\mu} \delta_{\alpha\nu} \delta^4(x-y) - 
\partial^y_\beta \partial^x_\mu {\cal K}_{\nu\alpha}(x,y)\right] 
G^{\rm S}_{\mu\nu}(x) 
\nonumber\\
&+& \int d^4x \,  \left[{1\over 2}(\partial \phi(x))^2 + V(\phi^2(x))\right], 
\label{action2}
\end{eqnarray}
where the propagator ${\cal K}_{\nu\alpha}$ was defined by Eq. (\ref{KK}).
Finally, integrating by parts, we obtain
\begin{equation}
S(C_\mu^{\rm cl},\phi^{\rm cl}) = 
{g^2\over 2} \int_{S(\Gamma)} d S_{\sigma\gamma}(v) 
\int_{S(\Gamma)} d S_{\lambda\rho}(u) 
{\cal G}_{\sigma\gamma\lambda\rho}(v,u) 
+ \int d^4x \, \left[{1\over 2}(\partial \phi(x))^2 + V(\phi^2(x))\right] 
\label{action3}
\end{equation}
where the tensor ${\cal G}_{\sigma\gamma\lambda\rho}$ is 
the same as
given by Eq. (\ref{GG}). 
Comparing with Eq. (\ref{svm}) we 
conclude that 
${\cal G}_{\sigma\gamma\lambda\rho}$ plays the same role in the dual 
theory as the two point correlator in the Stochastic Vacuum Model
if we neglect the contribution of the Higgs field to the action 
in (\ref{action3}) . 
In the London limit the contribution of the Higgs field to (\ref{action3}) 
vanishes and the identification is exact. 

In the general case we are considering here also the Higgs part 
gives a contribution to the non-perturbative dynamics. But let us neglect 
the dependence of the Higgs field, via the equations of motion, 
on the strings and take into account the contribution coming from the Higgs 
part as a finite contribution to the string tension. 
Then, also in the general case, ${\cal G}_{\sigma\gamma\lambda\rho}$ 
can be considered equivalent to the QCD two-point non-local condensate 
and in principle gives  information on the validity of the decomposition (\ref{decom}) 
and on the existence and the behaviour of the $D$ and $D_1$ functions.

Notice that in  the derivation of Eq. (\ref{action3}) we have not considered 
surface-like contributions which would arise from the functional 
integral on the right-hand side of Eq. (\ref{main}) once singular 
Higgs phase contributions are taken into account (these are 
also called  Abrikosov--Nielsen--Olesen strings). These surface 
terms would interfere with the surface terms coming from the external 
quarks loop. We make the assumption that these interference terms 
are unimportant in order to evaluate the long-range behaviour of the 
(heavy quark) Wilson loop average after the duality assumption (\ref{main}). 
In this way all the contributions coming from the singular Higgs phase 
factorize in the functional integral to a constant and 
play no role in the dynamics
(see also the discussion on this assumption made in the context of the 
London limit in Sec. 3).

We now evaluate (\ref{action3}) beyond the London limit.
 
Let us write a point $x$ in the four dimensional Minkowski space as 
$x = (x_{\|},x_\perp)$, where $x_{\|}=(x_1,x_4)$ and 
$x_\perp=(x_2,x_3)$ are now two-dimensional vectors. 
Let us indicate with small letters the components of $x$ 
belonging to $x_{\|}$ (e. g. $x_a$,$x_b$, ...) and with 
capital letters the components of $x$ belonging to $x_\perp$ 
(e. g. $x_A$,$x_B$, ...). In order to simplify the problem and to allow us 
to give an  analytic evaluation of (\ref{action3})
we choose the surface $S(\Gamma)$ (see Eq. (\ref{dirac})) 
to belong to the plane $x_\perp = 0$.  
It is reasonable in this case to assume, at least far away from the charge 
sources (i. e. in the middle of the flux tube), that the Higgs field 
depends only on the transverse coordinate $x_\perp$:
\begin{equation}
\phi = \phi(x_\perp).
\label{ass}
\end{equation}
We will make this crucial assumption for the rest of this section.

From Eqs. (\ref{ass}) and (\ref{KK}) we have 
${\cal K}_{\mu\alpha}(x,y)={\cal K}_{\mu\alpha}(x_\|-y_\|,x_\perp,y_\perp)$. 

In this situation we have that Eq. (\ref{action3}) can be written as:
\begin{eqnarray}
S(C_\mu^{\rm cl},\phi^{\rm cl}) &=&  
{g^2\over 2} \int_{S(\Gamma)} d S_{14}(x_\|) 
\int_{S(\Gamma)} d S_{14}(y_\|) 
{\cal G}_{1414}(x_\|-y_\|) + \hbox{Higgs sector},  
\label{action4}\\
{\cal G}_{1414}(x_\|-y_\|) &=&
\delta^4(x_\|-y_\|) - \epsilon_{14AB}\epsilon_{14CD} 
\partial^y_C \partial^x_A \left. {\cal K}_{BD}(x_\|-y_\|,x_\perp,y_\perp) 
\right|_{x_\perp=y_\perp=0}. 
\label{GG4}
\end{eqnarray}
After some simple manipulations it is possible to obtain from 
Eq. (\ref{KK}) an equation only for the transverse components of the gauge 
field propagator:
\begin{eqnarray}
&~& \left[ \partial^2_\perp \delta_{CB} - \partial_B\partial_C 
-  e^2 \phi^2(x_\perp) \delta_{CB} \right] 
{\cal K}_{CA}(x_\|-y_\|,x_\perp,y_\perp) 
\nonumber\\ 
&+&  \partial^2_\| \left[ \delta_{CB} - \partial_B(\partial_\perp^2 
-  e^2 \phi^2(x_\perp))^{-1}
\partial_C\right]{\cal K}_{CA}(x_\|-y_\|,x_\perp,y_\perp) = 
-\delta_{BA}\delta^4(x-y),
\label{KK4}
\end{eqnarray}
where $\partial^2_{\|} \equiv  \partial_a \partial_a$
and $\partial^2_{\perp} \equiv  \partial_A \partial_A$.

We look for a solution of Eq. (\ref{KK4}) of the type:
\begin{equation}
\epsilon_{14CD} \partial^y_C 
\left.{\cal K}_{BD}(x_\|-y_\|,x_\perp,y_\perp)\right|_{y_\perp=0} \equiv 
- \epsilon_{14CB} x_C {\cal K}(x_\|-y_\|,x_\perp).
\label{cho4}
\end{equation}
This is reasonable since in the transverse plane we have rotational invariance.
The function $\cal{K}$ is unknown, but from Eq. (\ref{KK4}) we have that 
it satisfies the equation:
\begin{equation}
(\partial^2 -  e^2 \phi^2(x_\perp) )x_A {\cal K}(x_\|,x_\perp)
= - \delta^2(x_\|) \partial_A \delta^2(x_\perp). 
\label{keq4}
\end{equation}

In the limit for $x_\perp \to 0$, we look for a solution  
$x_A {\cal K}$ of the type:
\begin{equation}
x_A {\cal K}(x_\|, x_\perp) \equiv \partial_A {\cal K}^{\rm p}(x)  
+ x_A f(x_\|) g(x_\perp), 
\label{ans}
\end{equation}
where ${\cal K}^{\rm p}$ is defined by 
\begin{eqnarray}
\partial^2 {\cal K}^{\rm p} &=& - \delta^4 (x) 
\nonumber \\
{\rm therefore} \, \, {\cal K}^{\rm p} &=& {1\over (2 \pi)^2} {1\over x^2}
\nonumber
\end{eqnarray}
and we normalize $f$ by imposing $\displaystyle \int d^2x_\| f(x_\|) =1$. 
The unknown functions $g$ and $f$ satisfy the equation:
\begin{eqnarray}
&~& \partial_\|^2 f(x_\|) \left( x_A g(x_\perp)\right) 
+ f(x_\|) \partial_\perp^2 \left( x_A g(x_\perp)\right) 
-  e^2 \phi^2(x_\perp)  f(x_\|)\left( x_A g(x_\perp)\right) = 
\nonumber\\
&~& \qquad\qquad\qquad 
 e^2 \phi^2(x_\perp) \partial_A {\cal K}^{\rm p}(x).
\label{fgeq}
\end{eqnarray}
Integrating over the longitudinal coordinates both side of the 
equation, we get 
$$
\left( \partial_\perp^2 -  e^2 \phi^2(x_\perp) \right) 
\left( x_A g(x_\perp)\right)
= - {1\over 2 \pi} e^2 \phi^2(x_\perp) {x_A \over x_\perp^2},
$$
where we used $\displaystyle\int d^2x_\| \partial_A {\cal K}^{\rm p}(x) 
= -\displaystyle{1\over 2\pi} {x_A\over x_\perp^2}$. 
This is exactly Eq. (\ref{cnpert2}) of the Appendix. Moreover, also 
the boundary  conditions are the same, since
$$
C_{\mu}(x) = \int d^4y \, {\cal K}_{\mu\alpha}(x,y) \,  
\partial_\nu G_{\nu\alpha}^{\rm S}(y).
$$ 
Therefore, a solution exists (for small $x_\perp$) and is given by
\begin{equation}
g(x_\perp) =  {e\over 2\pi} {C^{\rm np}(x_\perp) \over x_\perp}. 
\label{gperp}
\end{equation}
For the definition of $C^{\rm np}$ see the Appendix.
Using the expansion (\ref{Csmall}), for small $x_\perp$ we have:
\begin{eqnarray}
g(x_\perp) &=& {S_c\over 2\pi} - {S_\phi^2 \over 16 \pi} x_\perp^2 + \cdots,
\nonumber\\
x_A g(x_\perp) &=& {S_c\over 2\pi} x_A + \cdots, 
\nonumber\\
\partial_\perp^2 x_A g(x_\perp) &=& -{S_\phi^2\over 2\pi} x_A + \cdots,
\nonumber
\end{eqnarray}
where $S_c$ and $S_\phi$ are some constants defined as:
$S_c \equiv \displaystyle{\mathop{\lim}\limits_{x_\perp \to 0}}
e C^{\rm np}(x_\perp)/x_\perp$ and 
$S_\phi \equiv \displaystyle{\mathop{\lim}\limits_{x_\perp \to 0}}
e \phi(x_\perp)/x_\perp$. By solving numerically  the static 
equations of motions  (\ref{c2}) and (\ref{phi2}) (with quark sources 
at infinities) these constants can be calculated as a function of 
the Ginzburg--Landau parameter $\lambda/ e^2$, see Tab. 1, where 
for convenience we have introduced the 
dimensionless
quantities 
$S_c^\prime \equiv S_c / M^2$ and $S_\phi^\prime \equiv S_\phi / M^2$. 
The numerical solution of the equations of motion 
shows
that 
both $S_c$ and $S_\phi$ exist, are real and positive \cite{baknum}. 
Expanding Eq. (\ref{fgeq}) for small $x_\perp$ and keeping 
only the leading terms, we get an equation for 
the function
$f$:
\begin{equation}
\partial_\|^2 f(x_\|) =  { S_\phi^2 \over S_c} f(x_\|) 
- { S_\phi^2 \over S_c} \delta^2(x_\|).
\label{feq}
\end{equation}
A solution of this equation is:
\begin{equation}
f(x) = {1\over 2 \pi \ell^2} K_0\left({|x|\over \ell}\right),
\label{fsol}
\end{equation} 
where $\ell \equiv \displaystyle {\sqrt{S_c}\over S_\phi}$. 
We remember that $K_0(|x|/\ell) \sim - \gamma 
+ \log 2 - \log (|x|/\ell)$ in the short-range region ($|x| \to 0$) and 
$K_0(|x|/\ell) \sim {\displaystyle \sqrt{{\pi\over 2} {\ell \over |x|}}
e^{-|x|/\ell}}$ in the long-range region ($|x| \to \infty$).

Since  a solution exists our technical assumptions (\ref{cho4}) and 
(\ref{ans}) are self-consistent. 

Putting Eq. (\ref{cho4}) into Eq. (\ref{GG4}) we obtain:
\begin{equation}
{\cal G}_{1414}(x_\|-y_\|) = -\partial^2_\| {\cal K}^{\rm p} (x_\| -y_\|) 
+ {S_c\over \pi} f(x_\| -y_\|). 
\label{GG4sol}
\end{equation}
The long-distance exponential fall off and the 
weakly singular
($\sim \log(|x|)$) 
short range behaviour of the non-perturbative contribution to ${\cal G}_{1414}$ 
in Eq. (\ref{GG4sol}) is compatible with the lattice parameterization (\ref{para}). 
This fact provides an extremely interesting consistency check to the validity of 
the duality assumption (\ref{main}). Moreover this suggests the identification of  
the correlation length $T_g$, associated with the long-range behaviour 
of the QCD non-local condensate with the dual quantity $\ell$ 
(see  Eq. (\ref{fsol})). Notice that 
at variance with respect to the London limit result, here the correlation 
length is not simply given by the mass $M$ of the dual gluon.  

\begin{figure}[htb]
\vskip 0.8truecm
\makebox[0.2truecm]{\phantom b}
\epsfxsize=14.8truecm
\epsffile{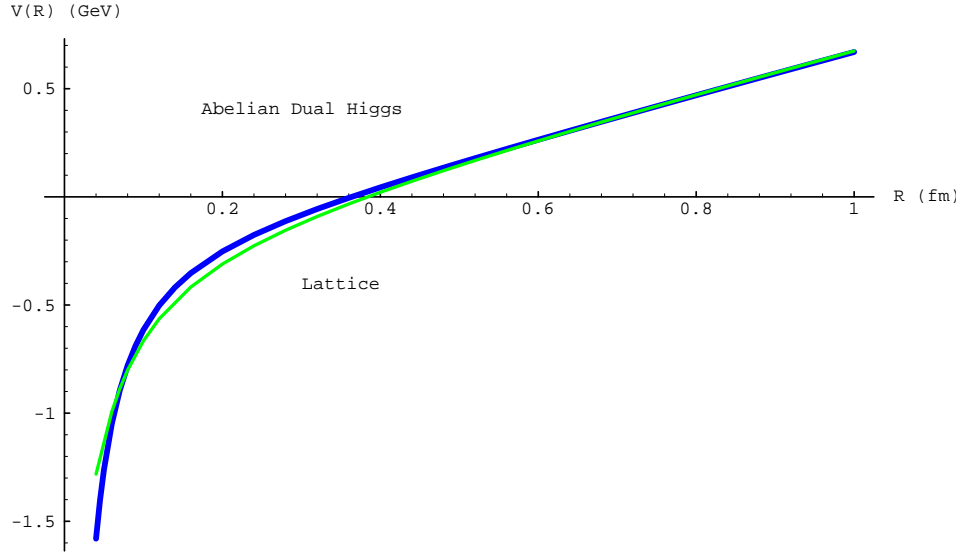}
\vskip -10truecm
\caption{\it The static potential of Eq. (\ref{V0dh}) for a superconductor 
on the border  between type I and type II with $\phi_0 = 210$ MeV, $\ell = 0.22$ fm 
and $\displaystyle{4 \over 3} {g^2\over  4 \pi}$ = 0.32 in comparison with the 
lattice fit of Ref. [30].} 
\label{figv0}
\end{figure}

Due to the almost regular short range behaviour of the non-perturbative 
part of Eq. (\ref{GG4sol}) the static potential can be calculated exactly 
without the use of an ultraviolet cut-off (at variance with respect to the 
London limit case, see Eq. (\ref{sigmaoo})), and it is given by 
\begin{eqnarray}
V_0(R) &=& {\mathop{\lim}\limits_{T \to \infty}} {1\over T}  
S(C_\mu^{\rm cl},\phi^{\rm cl})  
\nonumber\\
&=& {g^2 \over 2\pi} S_c \int_0^R dx_1 2 (R-x_1) \int_{-\infty}^{+\infty} dx_4 
{1\over 2\pi\ell^2} K_0\left({\sqrt{x_4^2+x_1^2}\over \ell}\right) 
- {g^2 \over 4\pi}{1\over R} 
\nonumber\\
&~& + \, {\rm Higgs} \, {\rm contributions}
\nonumber\\
&=& R \, {g^2 \over 2 \pi}S_c  + \left(e^{-R/\ell}-1\right){g^2 \over 2 \pi}S_c \,\ell 
- {g^2 \over 4\pi}{1\over R} + R \,\sigma_{\rm H} 
\label{V0dh}\\
&{\mathop{\longrightarrow}\limits_{ R \to \infty}}&  R\,{g^2 \over 2 \pi}S_c 
+ R \,\sigma_{\rm H}. 
\label{string}
\end{eqnarray}
For the simple case of $\phi = \phi(x_\perp)$ the Higgs contribution to the 
static potential turns out to be  given by a linear term with string 
tension $\sigma_{\rm H}$.\footnote{
The comparison between Eq. (\ref{V0dh}) and Eq. (\ref{v0svm})  
suggests to identify $D(x_\|)$ with $f(x_\|)S_c/\pi$. 
The same string tension (for what concerns the non-Higgs part) 
would, then, be obtained by  using equation (\ref{sigma}). 
We see, therefore, that the string tension is always emerging 
in the limit of large interquark distances and via an 
integral on a function depending on the correlation length.
Therefore our calculation confirms the existence of the non--local 
condensate and traces their origin back to a dual Meissner effect.} 
Taking explicitly into account this contribution, the total string 
tension is $\sigma= \displaystyle{g^2 \over 2 \pi}S_c+ \sigma_{\rm H}= 
\phi_0^2 ( 2 \pi S_c^\prime + \sigma_{\rm H}^\prime)$ where 
$\sigma_{\rm H}^\prime \equiv \sigma_{\rm H}/M^2$. For some values 
of $\sigma_{\rm H}^\prime$ see Tab \ref{tabv0}. In particular,  
for a superconductor on the border ($\lambda/e^2 = 1/2$)
from Tab. \ref{tabv0} we have $V_0(R)=  \pi \phi_0^2  \left( R + 
\ell\left(e^{-R/\ell}-1\right) \right) - \displaystyle{g^2 \over 4\pi}
{1\over R}$. In order to compare this potential with the heavy 
quark static potential we have to multiply it by the colour factor 
$4/3$. For a typical value of $\phi_0 \simeq 210$ MeV we get $\sigma = 
\displaystyle{4\over 3} \pi \phi_0^2 \simeq (430 \, {\rm MeV})^2$.
In Fig. \ref{figv0} we compare the static potential of Eq. (\ref{V0dh}) 
for a superconductor on the border between type I and type II  
for some typical values of the parameters with the lattice fit 
of Ref. \cite{latpot}.

One of the most interesting point is to relate the dimensional parameters 
$F_2$ and $T_g$ , 
the gluon condensate and the correlation length 
of QCD, to the dimensional parameters $\phi_0$ and $\ell$, 
the Higgs condensate and this characteristic length 
in the dual Abelian Higgs model. Our derivation identifies $\ell$ with the 
correlation length $T_g$ and eventually explains the existence of a 
finite correlation length in terms of an underlying dual Meissner 
effect that gives a mass to the dual field. In the dual theory \cite{bak} 
using  trace anomaly it is possible to relate the Higgs condensate 
to the gluon condensate, $F_2 \sim \lambda \phi_0^4$. Using the above 
value of $\phi_0$ and $\lambda/e^2 = 1/2$, 
one obtains for the gluon condensate the value found by \cite{svz}
$F_2 \simeq 0.013 \,{\rm Gev}^4$. This is how originally it was shown 
in DQCD that the QCD vacuum is compatible with a dual superconductor on the 
border between type I and II \cite{baker1}. Finally, we notice 
that in pure gluodynamics the 
lowest dimensional gauge and Lorentz-invariant
operator has dimension  
four 
its vacuum expectation value is
the gluon condensate.
In the dual model we have 
a relevant condensate, 
the Higgs condensate 
$\phi_0^2$, of dimension two. This could yield 
some interesting consequences in renormalon physics \cite{zak}.

Via numerical solution of the coupled equations for $C_\mu$ and $\phi$
and subsequent numerical interpolation, it is also possible to calculate 
the form of all semirelativistic $q\bar{q}$ potentials \cite{BBBZ,bak}.  
In principle, we could obtain an analytic solution  also for  the spin 
dependent and velocity dependent potentials following the same line of this 
paper. However, to this aim  the calculation of different components 
of the tensor $\cal{G}_{\sigma\gamma\lambda\rho}$ 
is necessary and some technical difficulties arise due to the fact
that the simple assumption (\ref{cho4}) is no longer valid. 
In the present situation we can try to gain some indications from the 
London limit result. Although, as we have seen, this is not the right limit 
in which to calculate the potentials, the qualitative long-range behaviour 
for the field-strength correlator is reasonable. In fact, in that limiting 
case it is possible to calculate the whole tensor 
$\cal{G}_{\sigma\gamma\lambda\rho}$ unambiguously  in terms of some 
functions $D$ and $D_1$ (see Eq. (\ref{GGoo})). Once we accept that in 
the presence of the quarks the short range behaviour of the Higgs field 
would regularize these functions on the flux-tube string, using the formulas 
of \cite{bvpot} we can express all the heavy-quark potentials 
in terms of integrals over these functions $D$ and $D_1$. 
Since these functions are reasonably compatible with the lattice 
fit (\ref{para}) this would explain the striking similarities in
the long-distance behaviour of the potentials obtained in DQCD and in
the Gaussian approximation of QCD \cite{bvpot}.

As a final remark, we notice that the flux tube structure between
two heavy quarks has been obtained in DQCD \cite{bakerflux} as well
as in the Gaussian approximation of QCD \cite{rued} and the result 
compare very favourably in both cases with the lattice calculation 
\cite{bali}. The profile of the longitudinal electric field, 
i. e. along the string between the quarks, as a function of the transversal 
distance from the string  is controlled by the  penetration length 
in one case and by the correlation length in the other. 

\section{CONCLUSION}

Under the assumption that the infrared behaviour of QCD is described 
by an effective Abelian Higgs model we have related the non-perturbative 
behaviour of the gauge invariant two-point field strength correlator 
$\langle g^2 F_{\mu\nu}(x)\, U(x,y)F_{\lambda\rho}(y)U(y,x)\rangle$ 
in QCD with the dual field propagator in the Abelian Higgs model of
infrared QCD. In this way the origin of the nonlocal gluon condensate is traced 
back to an underlying Meissner effect and the phenomenological 
relevance of the Gaussian approximation on the Wilson loop is 
understood as following from the classical approximation in the dual 
theory of long distance QCD. In particular the correlation length $T_g$ of QCD, 
which we know from direct lattice measurements, can be expressed completely 
in terms of the dual theory parameters ($\ell$). As a further check we  have 
calculated analytically the static potential and the string tension
which are 
quantities
 directly related to 
phenomenology. It turns out that 
the string tension is given by an integral over a function of the correlation 
length which can be identified with the non-local gluon condensate.
 There is no cutoff introduced in this calculation 
since it is not performed in the London
limit. 
We have shown that this  limit is   quite unphysical in 
the presence of sources and is valid only in the case of large 
distance from the chromoelectric string (which is different from large 
$q \bar{q}$ distances). Finally, these results shed some light  
also on the fact that the heavy quark potentials turn out to be equivalent 
in the SVM and in DQCD at large $q \bar{q}$ distances.

\section*{Acknowledgments}

The authors would like to thank D. V. Antonov for useful and enlightening 
conversations. One of us (M. B.) would like to thank  the members 
of the Institute of Theoretical Physics at Heidelberg for their 
kind hospitality during the period that  this work was begun.
He would also like to thank Matt Monahan for his very important 
contributions to this work.
Two of us (N. B. and A. V.) gratefully acknowledge 
the Alexander von Humboldt Foundation for the financial support, 
the perfect organization and the warm environment provided to the fellows.
 
\vfill\eject

\appendix
\section*{}

In this Appendix we study the equation of motion (\ref{c2}) 
in the presence of two static sources $g$ and $-g$ 
evolving from time $-T/2$ to time $T/2$ in the positions 
$R/2$ and $-R/2$ of the $x_1$ axes respectively. 
Therefore $x_\perp = (x_2,x_3)$. Under these conditions the Dirac string 
is given by:
\begin{equation}
G^{\rm S}_{\nu\mu}(x) = g \epsilon_{\nu\mu 14} \delta^2(x_\perp) 
(\theta(x_4+T/2) - \theta(x_4-T/2))(\theta(x_1+R/2) - \theta(x_1-R/2))
\label{dirstat}
\end{equation}

Defining $C_A(x_\perp) \equiv  \displaystyle{1\over RT} \int d^2 x_\| 
C_A(x) $, Eq. (\ref{c2}) can be written as:
$$
(\partial_\perp^2 \delta_{AB} -\partial_A\partial_B 
-  e^2 \phi^2(x_\perp)\delta_{AB})C_B(x_\perp) = 
- g \epsilon_{BA14}\partial_B \delta^2(x_\perp).
$$
It is convenient to split the field into the sum 
of two parts, $C_A = C_A^{\rm p} + C_A^{\rm np}$, satisfying 
the equations:
\begin{eqnarray}
\partial_\perp^2 C_A^{\rm p}(x_\perp) &=& 
- g \epsilon_{BA14}\partial_B\delta^2(x_\perp), 
\label{cpert}\\
(\partial_\perp^2 \delta_{AB} -\partial_A\partial_B 
-  e^2 \phi^2(x_\perp)\delta_{AB})C_B^{\rm np}(x_\perp) &=& 
   e^2 \phi^2(x_\perp) C_A^{\rm p}(x_\perp).
\label{cnpert}
\end{eqnarray}
The solution of Eq. (\ref{cpert}) is 
\begin{equation}
C_A^{\rm p}(x_\perp) = -{1\over e}{\epsilon_{BA14}x_B \over x_\perp^2} 
=-{1\over e}{1\over x_\perp} \hat{\theta},
\label{cpertb}
\end{equation}
where we have used the Dirac quantization condition, $g = 2\pi/e$, and 
$\hat{\theta} \equiv (-x^3/x_\perp, x^2/x_\perp)$ is the angular unit vector 
in the transverse plane. Substituting (\ref{cpertb}) in (\ref{cnpert}) 
and defining $C_A^{\rm np}(x_\perp) \equiv \epsilon_{BA14} 
C^{\rm np}(x_\perp) x_B/x_\perp$ (or, which is the same,  
${\vec{C}}^{\rm np}(x_\perp) = C^{\rm np}(x_\perp) \hat{\theta}$, 
where ${\vec{C}}^{\rm np}$ is a vector in the transverse plane), we obtain 
\begin{equation}
\left( \partial_\perp^2 -  e^2 \phi^2(x_\perp) \right) 
\left( {x_A\over x_\perp} C^{\rm np}(x_\perp)\right)
= -  e \phi^2(x_\perp) {x_A \over x_\perp^2},
\label{cnpert2}
\end{equation}
or 
\begin{equation}
{d\over d x_\perp} 
\left({1\over x_\perp}{d\over d x_\perp}\left(x_\perp\,C^{\rm np}(x_\perp)
\right)\right) = e^2 \left( C^{\rm np}(x_\perp) - {1\over e x_\perp} \right) 
\phi^2(x_\perp).
\label{Cperp}
\end{equation}

We can solve equation (\ref{Cperp}) for small values of $x_\perp$, 
assuming $\phi(x_\perp)= \displaystyle{S_\phi x_\perp \over e}+ \cdots$,  
obtaining:
\begin{equation}
C^{\rm np}(x_\perp) = {S_c \over e}x_\perp - {S_\phi^2 \over 8  e}
x_\perp^3 + \cdots,
\label{Csmall}
\end{equation}
where $S_c$ and $S_\phi$ are some constants.

\vfill
\eject

\begin{table}
\vspace{2.5in}
\begin{tabular}{|l|l|l|l|l|}      
Type of Superconductor                      & 
$\lambda/e^2$          \qquad               &
$S_c^\prime$           \qquad \qquad \qquad &
$S_\phi^\prime$        \qquad \qquad \qquad & 
$\sigma_{\rm H}^\prime$\qquad \qquad \qquad \\ 
\hline
I                             & $1/32$ & 0.1125 & 0.2516 & 1.142    \\
{\rm between} I {\rm and } II & $1/2$  & 0.25   & 0.6    & $\pi/ 2$ \\
II                            & 2      & 0.38   & 1.017  & 1.82     \\
II                            & 8      & 0.568  & 1.823  & 2.06     \\
II                            & 16     & 0.685  & 2.49   & 2.16     \\
\hline
\end{tabular}
\vskip 1truecm
\caption{\it Some values of the dimensionless quantities 
$S_c^\prime$, $S_\phi^\prime$ and $\sigma_{\rm H}^\prime$ as a function 
of the Ginzburg--Landau parameter $\lambda/e^2$, obtained by solving 
the static equations of motions with quark sources at infinities.}
\label{tabv0}
\end{table}

\end{document}